\documentclass[%
 aip,
 jcp,
 amsmath,amssymb,
 reprint
]{revtex4-1}

\usepackage{graphicx}
\usepackage[usenames,dvipsnames]{color}
\usepackage{algorithm}
\usepackage{algpseudocode}

\begin{document}

\preprint{AIP/123-QED}
\title{How chromophore labels shape the structure and dynamics of a peptide hydrogel}
\author{Frederick Heinz}
\affiliation{ 
Freie Universität Berlin, Department of Biology, Chemistry and Pharmacy, Arnimallee 22, 14195 Berlin}

\author{Jonas Proksch}
\affiliation{ 
Freie Universität Berlin, Department of Biology, Chemistry and Pharmacy, Arnimallee 22, 14195 Berlin}

\author{Robert F. Schmidt}
\affiliation{Stranski-Laboratorium für Physikalische und Theoretische Chemie, Institut für Chemie, Technische Universität Berlin, Straße des 17. Juni 124,
10623 Berlin}

\author{Beate Koksch}
\affiliation{ 
Freie Universität Berlin, Department of Biology, Chemistry and Pharmacy, Arnimallee 22, 14195 Berlin}

\author{Michael Gradzielski}
\affiliation{Stranski-Laboratorium für Physikalische und Theoretische Chemie, Institut für Chemie, Technische Universität Berlin, Straße des 17. Juni 124,
10623 Berlin}

\author{Bettina G.~Keller}%
\email{bettina.keller@fu-berlin.de}
\affiliation{ 
Freie Universität Berlin, Department of Biology, Chemistry and Pharmacy, Arnimallee 22, 14195 Berlin}

\date{\today}
%\keywords{Hydrogels, Coiled-Coil, Persistence Length, Diffusion coefficient}

%----------------------------------------------------------------------------
\begin{abstract}
Biocompatible and functionalizable hydrogels  have a wide range of (potential) medicinal applications.  
In contrast to conventional hydrogels formed by interconnected or interlocked polymer chains, self-assembled hydrogels form from small building blocks like short peptide chains. This has the advantage that the building blocks can be functionalized separately and then mixed to obtain the desired properties.
However, the hydrogelation process for these systems, especially those with very low polymer weight percentage ($<1$~wt\%), is not well understood, and therefore it is hard to predict whether a given molecular building block will self-assemble into a hydrogel.
This severely hinders the rational design of self-assembled hydrogels.
In this study, we demonstrate the impact of an N-terminal chromophore label amino-benzoic acid on the self-assembly and rheology of hydrogel hFF03 (hydrogelating, fibril forming) using molecular dynamics simulations, which  self-assembles into $\alpha$-helical coiled-coils.
We find that the chromophore and even its specific regioisomers have a significant influence on the microscopic structure and dynamics of the self-assembled fibril, and on the macroscopic mechanical properties. 
This is because the chromophore influences the possible salt-bridges which form and stabilize the fibril formation. 
Furthermore we find that the solvation shell fibrils by itself cannot explain the viscoelasticity of hFF03 hydrogels.
% More general Context 1~2
Our atomistic model of the hFF03 fibril formation enables a more rational design of these hydrogels. 
In particular, altering the N-terminal chromophore emerges as a design strategy to tune the mechanic properties of these self-assembled peptide hydrogels. 
\end{abstract}
\maketitle

%----------------------------------------
%   I N T R O D U C T I O N
%----------------------------------------
\section{Introduction}
The applications of hydrogels \cite{kirschning2018chemical} as biomedical materials ranges from  drug delivery systems \cite{Laurano2021, Xing2017}, to tissue engineering\cite{VanVlierberghe2011} and wound dressing. 
Additionally, hydrogels whose viscoelastic properties can be controlled and which can be systematically functionalized can mimic the extracellular matrix \cite{Hellmund2019, nicolas20203d} or the glycocalyx, i.e. the glycoprotein coat on epithelial cells, and are are valuable tools for \textsl{in-vitro} cell cultures.
Classical hydrogels often consist of long polymers, which are cross-linked by covalent bonds or physical interactions. 
However, one may also have hydrogels obtained by self-assembly, for instance consisting of small molecular blocks that self-assemble into fibrils. 
In that context, peptides are particularly versatile molecular building blocks for self-assembled hydrogels \cite{dasgupta2013peptide}.
Despite the fact that the network structure in self-assembled hydrogels is stabilized by relatively weak hydrophobic or electrostatic contacts, these materials can retain an astonishingly high amount of water.
Often more than 99~wt\% (mass fraction) of the hydrogel is water \cite{Galler}.
While the traditional model of intertwined polymer chains can explain water retention\cite{Dargaville2022} and mechanical properties for a hydrogel with a high polymer mass fraction, this becomes less obvious to rationalize for self-assembled hydrogels with low polymer mass fraction.
Currently, our understanding of the molecular structure of self-assembled hydrogels is limited, because many traditional structure analysis methods, such as NMR or X-ray scattering face challenges in properly resolving the structure of very flexible and dynamic hydrogel networks.  The use of small-angle scattering to study such systems has been reviewed recently\cite{Mcdowall2022}.
As a result, it is hard to predict, whether a given molecule will form a hydrogel or not. 
Even small and inconspicuous changes to the protein structure or pH value can make or break a hydrogel.\cite{Li21}
Systematically designing the viscoelastic properties of a hydrogel is currently not feasible, because it requires a detailed understanding of the structure and dynamics of the hydrogel network. 

Recently, the coiled-coil-based peptide hFF03 (\textsl{h}ydrogelating, \textsl{f}ibril \textsl{f}orming) has been proposed as a scaffold for a functionalizable, biocompatible hydrogel \cite{Hellmund2021}.
hFF03 is a 26-residue peptide, which is designed to self-assemble into $\alpha$-helical coiled-coil dimers. 
The dimers are stabilized by a leucine zipper motif and exhibit several solvent exposed lysine residues to which functional groups, such as carbohydrates, can be attached. 
hFF03 self-assembles into a 3D fibril network\cite{Hellmund2019a} and has the viscoelastic properties of a hydrogel. 
The hydrogel nature of this peptide was verified with the tube inversion test at 4 wt\% (4\% polymer mass fraction), and the viscoelastic properties were determined by rheological experiments \cite{Hellmund2019a}. 
Fibril diameter and persistence length were determined by small angle neutron scattering (SANS) experiments\cite{Hellmund2019a}.
Furthermore, hFF03 retains its hydrogel character when functionalized with a carbohydrate moiety \cite{Hellmund2019a}.
However, the SANS data do not yield any insight into the structure of the fibrils at the molecular level and into the self-assembly mechanism.
Two mechanisms are possible for hFF03. 
First, the coiled-coil dimers could form such that the two peptides are aligned with zero lateral shift, leading to an aggregation via the charged termini of the coiled-coil dimers. 
Second, because the sequence of hFF03 consists of three heptad repeat units and a five-residue C-terminal segment, a sticky-end assembly \cite{pandya2000sticky}  is conceivable, where the C-terminal segment bridges the gap to the next coiled-coil dimer. 
In this mechanism, the fibrils are stabilized mainly by a hydrophobic interaction. 
Because of the bridge between consecutive coiled-coil dimers, this mechanism would immediately explain how long fibrils can arise.
We also made the intriguing observation that the presence and isomeric structure of the chromophore label aminobenzoic acid, which was coupled to the hFF03 peptide as a UV-Vis marker in Ref.~\citenum{Hellmund2019a}, has a drastic effect on the viscoelastic properties of the substance. 
It seems plausible that this is a result of the chromophore label interfering with the self-assembly mechanism. 
The purpose of this study is to construct an atomistic model of hFF03 which is consistent with the available structural data. 
Starting from this model, we will conduct MD simulations to elucidate the self-assembly mechanism and to understand how the presence of a chromophore label influences the fibril formation.
The goal is to identify the crucial microscopic interactions that determine the viscoelastic properties of hFF03 hydrogels.
%
%------------------------------
%
%
\section{Methods and system}
\label{sec:methods}
\subsection{System}
\begin{figure}[h!]
    \centering
\includegraphics[width=\columnwidth]{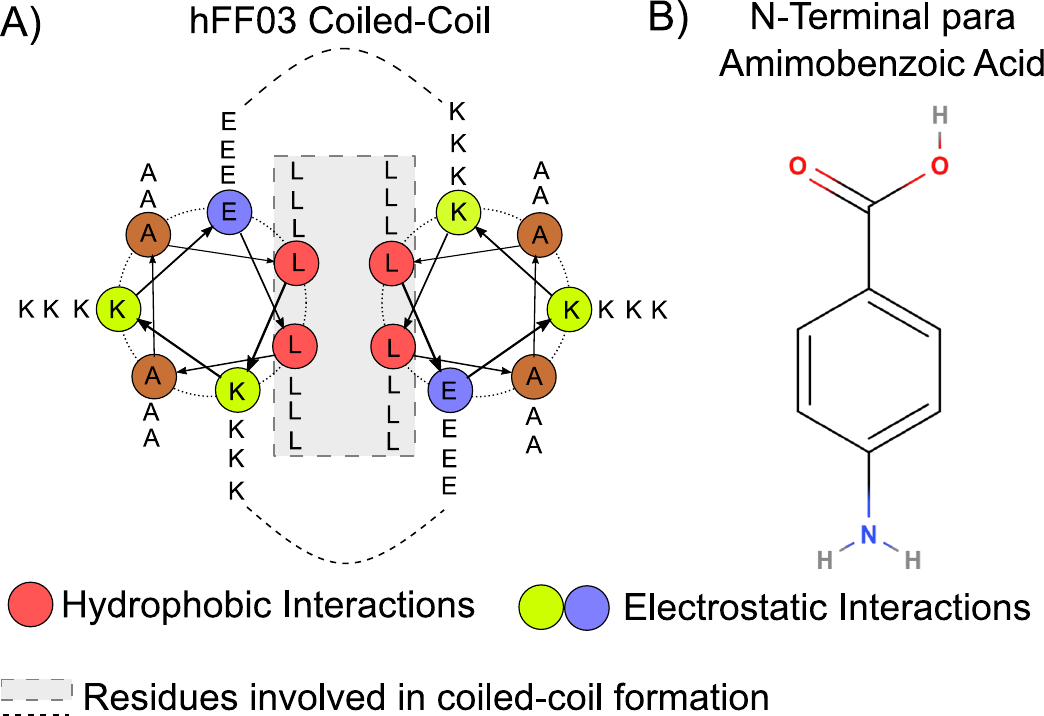}
    \caption{A) Coiled-coil structure for hFF03 proposed in Ref.~\citenum{Hellmund2021}. 
    The coiled coil is stabilized by salt bridges between the polar side chains as well as hydrophobic core consisting of a leucine zipper motif.
    B) N-terminal chromophore \textsl{para}-aminobenzoic acid. }
    \label{fig:hFF03}
\end{figure}

Each hFF03 peptide monomer consists of the sequence x-LKKELAA-LKKELAA-LKKELAA-LKKEL from N- to C-terminus (Fig.~\ref{fig:hFF03}.A).
"x" denotes an optional aminobenzoic acid (aba) group (Fig.~\ref{fig:hFF03}.B), which is covalently attached to the peptide  via a peptide bond between the carboxyl group of aba and the amino group of the peptide N-terminus.
Since aba is a chromophore, it allows for the convenient determination of peptide concentration through absorbance measurements.
In Ref.~\citenum{Hellmund2019a}, \textit{ortho}-aminobenzoic acid  was used as a chromophore label. 
We replicate and expand these experiments by synthesizing 
hFF03 without chromophore (no-hFF03), 
hFF03 with $x= $\textit{ortho}-aminobenzoic acid (oaba-hFF03), and
hFF03 with $x= $\textit{para}-aminobenzoic acid (paba-hFF03).

\subsection{Oscillatory Shear Rheology}
The rheological measurements were performed on an Anton Paar MCR 502 WESP temperature controlled rheometer in strain-imposed mode at physiological temperature (37°C). A chromium oxide-coated cone-plate measurement system was used with a diameter of 25\,mm, a cone truncation (gap width) of 48\,\textmu m and a cone angle of 1°. The temperature was set using a Peltier measuring system combined with a Peltier hood to ensure a minimized temperature gradient throughout the sample. To minimize evaporation, a solvent trap was used. The oscillation frequency was varied between 0.05 and 50\,Hz at a constant strain amplitude of 5\% (a preliminary amplitude sweep showed that this value is still in the linear viscoelastic regime). An up-- and down--sweep was performed to check for possible hysteresis effects and the results shown represent averages of both sweeps.

\subsection{Circular Dichroism Spectroscopy}
The oaba-hFF03 peptide was dissolved in Milli-Q water at three different concentrations and the pH adjusted to 7.4 with 1 M aqueous NaOH and HCl.
The obtained solutions were measured at 37 $^{\circ}$C 2 h after preparation using a Jasco J-810 spectropolarimeter (JASCO Deutschland GmbH, Pfungstadt, Germany) with a Jasco PTC-432S Peltier temperature element. Spectra were recorded using detachable Quartz Suprasil cuvettes with a path length of 0.1 mm (Hellma Analytics, Müllheim, Germany). Spectra are the mean of three measurements and background corrected by subtraction of a solvent spectrum.  

\subsection{Molecular-dynamics simulations}
For the peptide monomers we use the sequence LKKELAA-LKKELAA-LKKELAA-LKKEL. The initial structures of the coiled-coil dimers are created using the the web tool CCBuilder2.0\cite{Wood2018} by Derek Woolfson. Lysine residues and N-terminus are protonated (charge +1), glutamate residues and the C-terminus are deprotonated (charge -1), corresponding to the expected protonation at pH7. 
For oaba-hFF03 and paba-hFF03, the aminobenzoic acid group was added manually with Pymol\cite{pymol} and the force field was calculated with AmberTools \cite{Case2005}, with the gas charge calculation method. The amino group was protonated. A single coiled-coil dimer has a total charge of +8.\\
MD simulations were carried out with Gromacs2021+CUDA on the Curta cluster system\cite{Bennett2020}  with the Amber99SB-ILDN force field\cite{Lindorff-Larsen2010}. After energy minimisation and relaxation in NVT and NPT ensemble, the simulations are carried out in the NPT-ensemble. The temperature was maintained at $T=300~K$ using the velocity-rescale thermostat with a coupling constant of 0.1~ps. The pressure was maintained at 1 bar using the Parinello-Rahman barostat with a coupling constant of 2~ps. The simulation uses a leap frog integrator with 2~fs per step and and periodic boundary conditions in all three spatial direction. Covalant bonds to hydrogen atoms were constraint using the LINCS algorithm. 
Coordinates were written to file every 20~ps.
\\
Specific simulation set-ups are described below.
The force-field and all input files for the simulations are available via the code repository.

\paragraph{Oligomerisation state.}
Dimer and tetramer coiled-coil starting structures were created with CCBuilder2.0.
Each of these starting structures was solvated in TIP3P  water\cite{tip3p} and the simulation boxes were neutralized with 8 Cl$^-$ (dimers) or 16 Cl$^-$. 
After equilibration, the systems were simulated for 100~ns.
The diameters were evaluated from the last 50~ns of the simulation, specifically
$d(\mathrm{LEU-C}_{\alpha})$ = distance between opposite leucine C$_{\alpha}$-atoms,
$d(\mathrm{LYS-C}_{\alpha})$ = distance between opposite lysine C$_{\alpha}$-atoms,
$d(\mathrm{LYS-NH3^+})$ = distance between opposite lysine side-chain amino groups, measured at the $N$-atom.
See Fig.~\ref{fig:Oligomerisation}

\paragraph{Coiled-coil oligomers with lateral shift.}
A structure of a continuous coiled-coil with 16 LKKELAA heptad repeats was created with CCBuilder2.0\cite{Wood2018}, which served as the starting structure for the model CC in Fig.~\ref{fig:Lp}. To obtain starting structures for the models A, B and C, heptat repeat units were cut out from the CC structure and termini were fixed with pdbfixer \cite{pdbfixer}. 
Each starting structure was solvated with TIP3P  water\cite{tip3p} in a rectangular box with 2~nm space around the peptide in all directions. The boxes were neutralized with 32 Cl$^-$ anions for the models A,B and C and 34 Cl$^-$ for continuous coiled-coil. 
Five independent simulations of 150~ns were conducted for each system, but only the last 50~ns were used to calculate the persistence length. 

\paragraph{Self-assembled coiled-coils.} 
32 coiled-coil dimers of no-hFF03 , oaba-hFF03 or paba-hFF03 were solvated in TiP3P water in a cubic box with 20 nm box length, corresponding to roughly a 4\% polymer mass fraction. 256 Cl$^-$ anions were added to generate a neutral simulation box.
Three independent simulations of 150~ns were conducted for each of the systems.

\paragraph{Local diffusion coefficient.}
A single no-hFF03 coiled-coil dimer was solvated in TIP3P water in a simulation box with size 7~nm $\times$ 7~nm $\times$ 4.7~nm.
The box was neutralized with 8 Cl$^-$ anions and simulated for 0.2~ns with at simulation timestep of 1~fs. 
Coordinates were written to file every 10~fs.

The bulk water values for TIP3P were calculated from a simulation of pure TIP3P water with similar set-up and simulation time.

\subsection{Analysis of the MD simulations}
\paragraph{Persistence length.}
The persistence length is defined via the following correlation function  \cite{Boal2012}
\begin{equation}
    L_P(l)=\int_0^{L_{\mathrm{total}}}\langle\vec{e}\left(l\right)\vec{e}\left(l+ \Delta l\right)\rangle \, d\Delta l \, ,  
    \label{eq:LP1}
\end{equation}
where $\langle \dots \rangle$ represents the ensemble average, $\vec{e}\left(l\right)$ and $\vec{e}\left(l+ \Delta l\right)$ are unit vectors of the of the chain orientation at position $l$ and $l+\Delta l$, and $L_{\mathrm{total}}$ is the total chain length. 
Note that $L_P(l)$ depends on the initial point $l$.
For a chain with $N$ discrete beads, spaced at a distance of $\Delta l$, eq.~\ref{eq:LP1} simplifies to
\begin{equation}
    L_P(l)=\sum_{j=1}^{N}\langle\vec{e}_l\vec{e}_j  \rangle \cdot \Delta l  \, ,
    \label{eq:LP2}
\end{equation}
where $\vec{e}_j =(\vec{r}_{j+1} - \vec{r}_{j})/\Delta l$ is the local chain orientation at the $j$th bead, and $\vec{r}_{j+1}$ and $\vec{r}_{j}$ are the positions of bead $j$ and bead $j+1$.
See Fig.~\ref{fig:Lp} A for sketch.
In principle, the persistence length can be directly calculated from equation \ref{eq:LP2}. 
However, by construction ($\langle\vec{e}_l\vec{e}_j  \rangle \le 1$), eq.~\ref{eq:LP2} only yields values $L_P(l)$ that are lower than the chain length $L_{\mathrm{total}} = (N-1) \Delta l$. 
For short and stiff chains this is unphysical.
Under the assumption that a more bent chain is in an higher energy state, it is possible to calculate the expected deformation of a chain with the Maxwell Boltzmann relation and derive the following equation\cite{Boal2012}:
\begin{equation}\langle\vec{e}\left(l\right)\vec{e}\left(l+ \Delta l\right)\rangle = e^{-\Delta l/L_P}\, ,
    \label{eq:LP3}
\end{equation}
i.e.~the correlation function decays exponentially with the distance from $\vec{r}_1$.
For each coiled-coil chain in Fig.~\ref{fig:Lp}, we define a chain of beads with one bead per leucine-leucine zipper pair, where the bead
position is the midpoint between the leucine C$_{\alpha}$ atoms.
We estimated the left-hand side of eq.~\ref{eq:LP3} and fitted an exponential function to obtain $L_P$.
The persistence length in the simulations of self-assembled coiled-coil oligomers were obtained by the same approach, where the oligomer chains were identified by the approach described below.

\paragraph{Identification of oligomer chains.}
Continuous coiled-coil oligomer chains in our simulations of self-assembled  coiled-coil oligomers were identified based on salt bridges between coiled-coil dimers. 
A contact between an amonium group and a carboxyl group counts as a salt bridge, if a hydrogen of the NH$_3^+$ group is within 0.35~nm of either oxygen of the COO$^-$ group.
We defined a dimer-dimer interaction as a salt bridge between the N-terminus, K2 or K3 of one dimer and the C-terminus or E25 of the neighboring dimer.
Our code detects self-assembled chains and in case of a chain split, treats both chains as individual chains. It also detects if the chain binds to itself and forms a circular fibril chain. In this case it stops the chain length and a circular chain is treated as one chain element longer than the sum of its chain elements. 

\paragraph{Local self diffusion coefficient.}
The diffusion constant $D$ of a particle is related to its velocity autocorrelation function (VACF) via
\begin{eqnarray}
    D &=&\int_{\tau=0}^{\infty} \langle v(0)v(\tau)\rangle\, \mathrm{d}\tau
\label{eq:D}    
\end{eqnarray}
(Green-Kubo relation), where $v(0)$ is the velocity at time $t=0$ along a single spatial coordinate,  $v(\tau)$ is the velocity at time $\tau$, and $\langle \dots \rangle$ denotes an ensemble average.
To calculate the velocity  auto-correlation function VACF $ \langle v(0)v(\tau)\rangle$, we use the Wiener-Kinchin\cite{RADER1967} theorem.
This allows for the direct calculation of the VACF from the velocities $v(t)$ with the help of the Fast Fourier Transformation algorithm FFT\cite{FFT}. 
The theorem can be summarized such that the VACF is the real part of the inverse FFT of the FFT multiplied by its complex conjugate. 
\begin{equation}
     \langle v(0)v(\tau)\rangle= ir\mathrm{FFT}(~\mathrm{FFT}(v(t))*\mathrm{FFT}(v(t))~)
     \label{eq:VACF}
\end{equation}
To obtain the non-circular autocorrelation function and prevent problems with odd lengths, it is necessary to use a so-called zero padding\cite{Hilbert2013}.

To obtain a local water diffusion coefficient dependent on the radial distance to the coiled-coil, we use a bead chain, as defined before, to locate the center of the coiled-coil.
The coiled-coil dimer axis is the vector between the first and last bead of the coiled-coil. 
The radial distance is the distance of the center-of-mass of a water molecule to this axis. 
Algorithm 1 summarizes our approach.

\begin{algorithm}[H]
\caption{Local Diffusion Coefficient}
\label{alg:Diff}
\begin{algorithmic}
\State Discretize the radial distance into bins
\For{each H$_2$O in a trajectory}
    \State Calculate  $\vec{v}_{COM}$ \Comment{\textit{COM velocity}}
    \State Calculate  $\vec{q}_{COM}$ \Comment{\textit{COM position}}
    \State Cut trajectory into 5 ps long segments
    \For{each trajectory segment} \Comment{\textit{calculation of} VACF}
        \State Assign trajectory segment to radial distance bin using $\vec{q}_{COM}$
        \State calculate VACF via FFT (including zero padding)
    \EndFor
    \State Add the VACF in each distance bin and normalize 
\EndFor
\For{each bin}
    \State Normalize by the number of VACFs.
    \State Numerically integrate VACF (eq.~\ref{eq:D})
\EndFor
\end{algorithmic}
\end{algorithm}

The FFT is carried out for each spatial direction of $\vec{v}_{COM}$ separately. To calculate the VACF via FFT, proceed as follows: 
\begin{itemize}
        \item Zero padding\cite{Hilbert2013}
        \item $FFT(\bar{v}_{COM})$
        \item Multiply by its complex conjugate
        \item $ir$FFT (real part of inverse FFT)
        \item Reverse zero padding
        \item Divide by number of frames. (This step depends on the used FFT implementation)
\end{itemize}
Then average over all three dimensions.

\subsection{Code Repository}
All code for the analysis of the trajectories and their visualisation  is written in Julia1.7.3\cite{julia} and is accessible in the code repository with short example files.
\href{https://github.com/bkellerlab/hFF03_hydrogels}{https://github.com/bkellerlab/hFF03\_hydrogels}

%------------------------------------
%
%
\section{Results}

\subsection{Mechanical properties of hFF03-hydrogels}
We created probes following the experimental conditons of Hellmund et. al.\cite{Hellmund2021} and determined their viscoelastic properties using oscillatory shear rheology.
In an oscillatory shear experiment, the material is subjected to an oscillating shear strain and the resulting shear stress is  measured. The ratio of stress and strain yields the complex modulus $G^*=G'+iG''$, which is a measure of the material's overall resistance to deformation. Here, $G'$ is the storage modulus, representing the elastic contribution and $G''$ is the loss modulus, representing the viscous contribution. In a frequency sweep experiment, the oscillation frequency is varied while the strain amplitude is kept constant. $G'$ and $G''$ are measured as a function of the oscillation frequency. Their values and frequency dependence are characteristic of the linear viscoelastic properties of a material.
In hydrogels, the elastic contribution (storage modulus $G'$) is usually larger than the viscous contribution (loss modulus $G''$), i.e. $G' > G''$, and both moduli are rather insensitive to frequency \cite{Picout2003}.
In an ideal system with fibrils that are cross-linked by strong interactions of covalent bonds, both lines would be parallel. But real systems often deviate from this ideal behaviour and show more complex behaviour\cite{Garrec2012}. 
\begin{figure}
    \centering
    \includegraphics[width=\columnwidth]{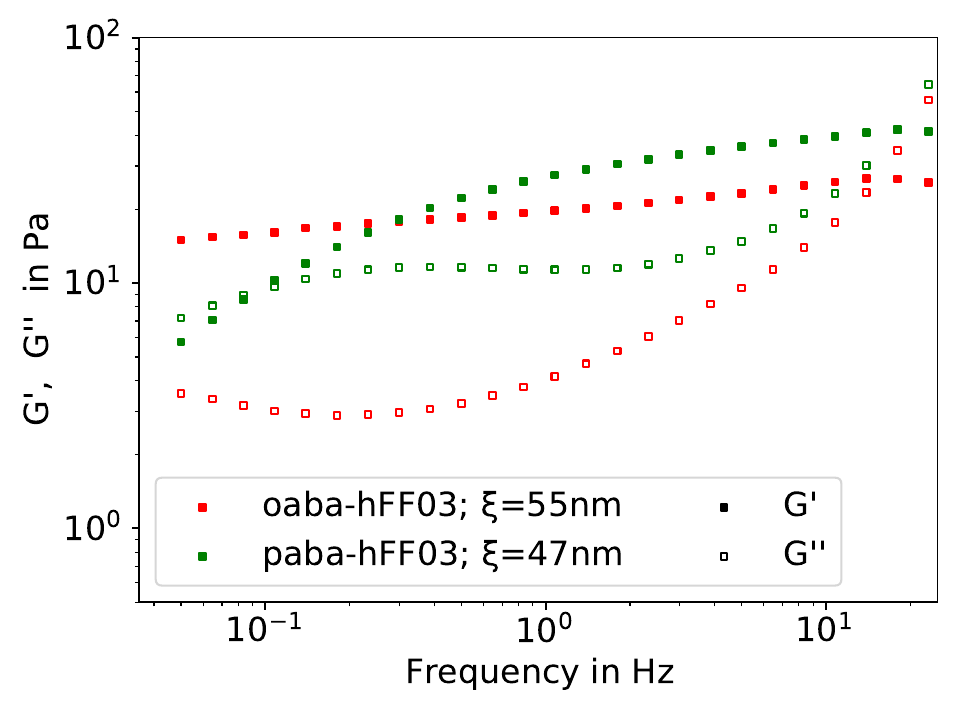}
    \caption{Frequency sweep measured through oscillatory shear experiment of hFF03 variants. 
    $G'$: storage modulus; $G''$: loss modulus;  $\xi$: mesh size, evaluated at 8.34 Hz.}
    \label{fig:Rheo}
\end{figure}
While it is generally assumed that a chromophore has only little influence on the macroscopic properties of a system, we found that varying the chromophore induced striking differences in the viscoelastic propeties of hFF03 (FIG.~\ref{fig:Rheo}).
oaba-hFF03 and paba-hFF03 exhibit typical hydrogel behaviour, while no-hFF03 shows no such behaviour and is a low viscous liquid. The moduli of the hydrogels have values in the range of 5-40 Pa and their behaviour is distinctly gel-like, since $G' > G''$. Their gel-character is also corroborated by the tube inversion test. While both isomers exhibit gel-like behaviour, oaba-hFF03 has a larger difference between $G'$ and $G''$ and it is clearly the stronger hydrogel of the two, as here also $G'$ is rather constant.  In contrast, for paba-hFF03 $G'$ becomes substantially smaller at lower frequency, which means that its structural relaxation time is much shorter and it will relax mechanical stress after much shorter times than oaba-hFF03. 
We can use the value of $G'$ in the plateau region, $G_0$, to give an estimate for the characteristic size $\xi$, i.e., the mesh size of the hydrogel network \cite{Gennes1979,Pincus1976,Tsuji2018}. Assuming a mesh of size $\xi$, where each mesh stores an energy of $k_BT$, one arrives at the relation: $\xi = (G_0/k_BT)^{-1/3}$. For the calculation of $\xi$, we take $G_0$ to be the value of $G'$ at a frequency of 8.34\,Hz. Using $G_0$, we find mesh sizes of around 55 and 48\,nm for oaba-hFF03 and paba-hFF03, respectively. Since no true plateau is seen for $G'$, the chosen frequency is rather arbitrary and the values for $\xi$ can be regarded as an upper estimate.

The viscoelasticity of no-hFF03 was too low to be accurately measured at higher frequencies and we present the results  in the SI (FIG. 1) solely for the sake of completeness.
We conclude that the chromophore is not an ``innocent'' label. Its presence is critical for hydrogel formation in hFF03, and even the position of the amino group in the chromophore influences the stability and mechanical properties of the hydrogel.
To better understand this behaviour, 
we set out to construct an atomistic model of self-assembled hFF03 peptides.

%-----------------------------------------------
%   M O D E L   B U I L D I N G
%-----------------------------------------------
\subsection{Structural model}
Previous SANS and cryogenic transmission electron microscopy (Cryo-TEM) showed that oaba-hFF03 self-assembles into fibrils with a diameter of 2.28~nm to 2.60~nm and a persistence length of 9 to 14~nm \cite{Hellmund2021}. 
Because of the design of the peptide sequence \cite{Zacco2015}\cite{Woolfson2017}, we can assume that the peptides form $\alpha$-helices which have a hydrophobic flank consisting of leucine residues. 
The peptides can self-assemble into coiled-coils via this hydrophobic region (leucine zipper motif). 
For the atomistic model, we need to consider the following aspects: 
($i$) oligomerization state (How many peptide strands are assembled across the diameter of the fibril?);
($ii$) orientation the helices within the fibril (Do all peptides have the same N-to-C-terminus orientation along the fibril?), and
($iii$) lateral shift of the $\alpha$-helices within respect to each other. 

\subsubsection{Oligomerization State and Helix Orientation}
\begin{figure}
        \centering\includegraphics[width=\columnwidth]{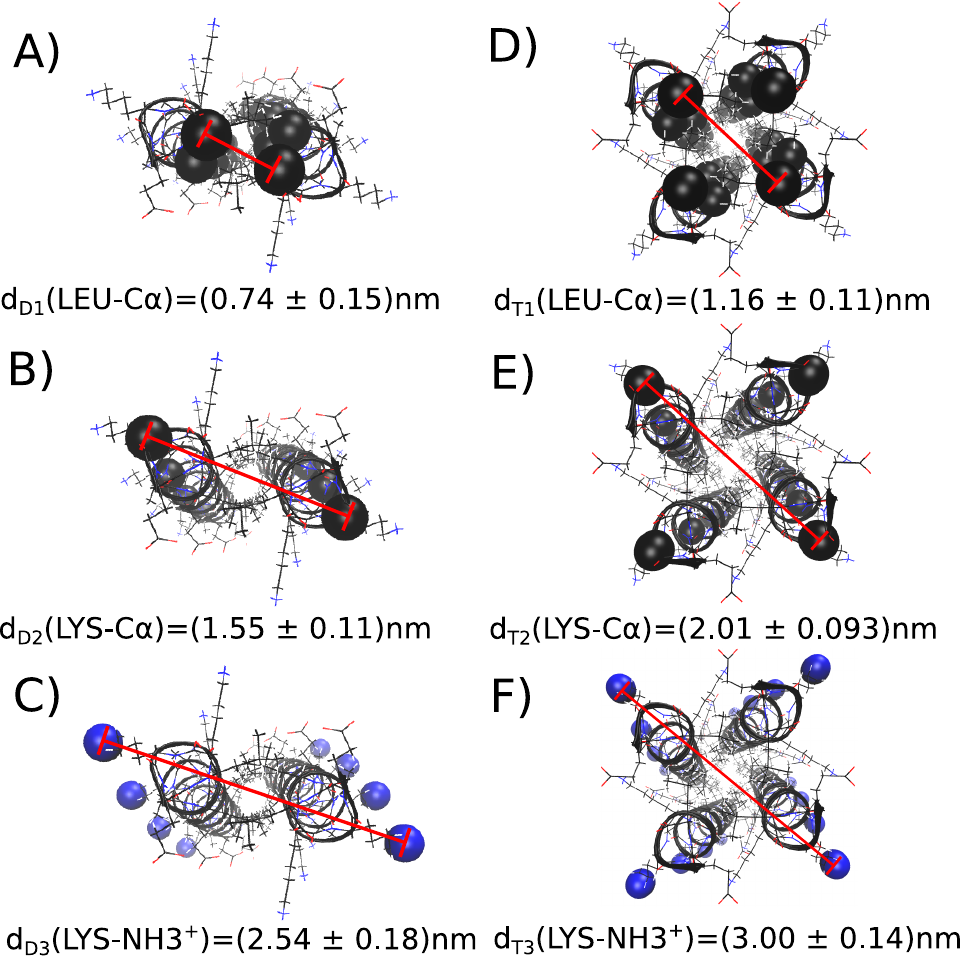}
        \caption{Diameter of coiled-coil dimers (\textbf{A},\textbf{B},\textbf{C}) and coiled-coil tetramers (\textbf{D},\textbf{E},\textbf{F}), averaged from   50~ns MD simulations for each model. Experimental fibril diameter from the SANS experiments range from 2.28 to 2.6~nm. \cite{Hellmund2021}}
        \label{fig:Oligomerisation}
\end{figure}
The oligomerization state is the number of peptide helices that self-assemble into a coiled-coil oligomer and the relative orientation of these helices. 
hFF03 is designed following the rules of coiled-coiled building formulated by Woolfson et al. \cite{Woolfson2005a}, and thus it is highly likely that hFF03 self-assembles into a parallel coiled-coil dimer. 
Sequences can be tested for their preferred oligomerization state using the web tool LOGICOIL\cite{Logicoil}. For hFF03, the preference is for parallel dimer with 0\% chance of forming a trimer.
This structure of a parallel dimer has also been proposed by Hellmund et al.\cite{Hellmund2021}. 
We used MD simulations to verify this model and simulated different oligorimerization states and helix orientations of no-hFF03.
The models were constructed without lateral shift, and we assume that the chromophore has no influence on the oligomerization state.
Coiled-coils modeled as anti-parallel dimer were not stable and drifted apart in the simulation. 
By contrast, coiled-coil models as parallel dimer and parallel tetramer remained stable during 50 ns MD simulation.
In the tetramer model, the leucine flanks of the four $\alpha$-helices form a joint hydrophobic core. 
By contrast, modelling a tetramer as  two parallel dimers next to each other did not yield a stable complex.
FIG.~\ref{fig:Oligomerisation} compares the diameter of the coiled-coils, measured at different reference points, to the experimental diameter\cite{Hellmund2021}.
Taking the amino groups of opposing lysine side chains as reference points (FIG.~\ref{fig:Oligomerisation}. C and F) likely best represents the SANS experiment. 
The diameter of the parallel dimer model (FIG.~\ref{fig:Oligomerisation}. C) matches the experimental results of 2.28 nm to 2.60 nm \cite{Hellmund2021}, while the diameter of the tetramer dimer model (FIG.~\ref{fig:Oligomerisation}. C) is too large.
Thus, the parallel dimer coiled-coils are consistent with the SANS experiment as well as the prediction by LOGICOIL\cite{Logicoil}, and we used this model in all further simulations.

\begin{figure*}
\includegraphics[width=16cm]{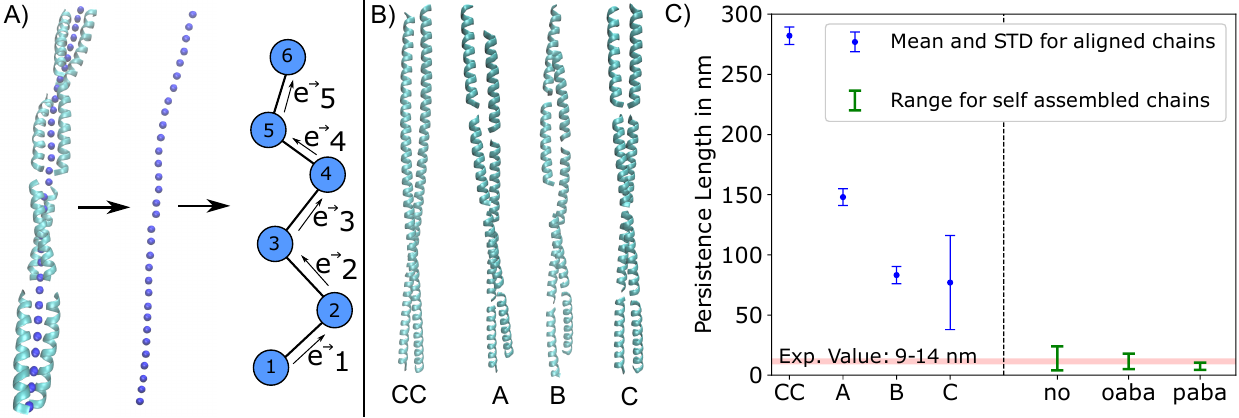}
    \caption{
    \textbf{A)} Virtual bead chain at the center of a coiled-coil fibril chain, which we used for the calculation of chain vectors $\vec{e}_i$ in eq.~\ref{eq:LP2}. 
    \textbf{B)} Models of coiled-coil fibril chains. 
    CC:  continuous coiled-coil with LKKELAA heptat repeats. 
    A: lateral shift, but still a continuous leucine zipper
    B: small and large lateral shift, the zipper is interrupted at the large lateral shift. 
    C: zero lateral shift, chain of individual coild-coil dimers.
    \textbf{C)} Persistence lengths for different models. CC, A, B and C are results for aligned fibrils, and hFF03, oaba-hFF03 and paba-hFF03 for self assembled fibril chains. Both codes have slightly different chain definitions and cannot be directly compared. The range for self assembled fibrils is the total range of observed values over all chain lengths with more than 2000 data points. The additionally constrained to a maximum 7 coiled-coils, since the longer chains behave too erratic.
    Experimental persistence length are from SANS measurements \cite{Hellmund2021}.
    }
    \label{fig:Lp}
\end{figure*}
\subsubsection{Lateral shift and persistence length}
Next, we discuss the lateral alignment of the two parallel $\alpha$-helices within the coiled-coil fibril.
Because of the repeated motif LKKELAA in the peptide sequence of hFF03, $\alpha$-helices could self-assemble such that the termini do not line up, but are shifted by one or two repeat motifs (lateral shift, sticky-end assembly) \cite{pandya2000sticky}.
hFF03 does not feature an anchor, such as a disulfide bridge, which could enforce a particular lateral shift. 
Sequences, for which lateral shift and overlapping helices have been reported, either have oligomerization states of five or more helices \cite{SA2001}, or consist of helices with different charges at either end \cite{Lou2019}. 

Importantly, fibrils with and without lateral shift self-assemble via different interfaces \cite{Bromley2011} and would result in different fibril flexibility.
With no lateral shift, each dimer can move independently of the other dimers, and elongated fibrils arise if the dimers align linearly, possibly stabilized by salt-bridges and hydrogen bonds between N- and C-termini of adjacent dimers (structure C in FIG.~\ref{fig:Lp}.). 
This would lead to very flexible fibrils, and it is unclear whether the resulting aggregate would be sufficiently stable to explain the observed viscoelasticity. 
By contrast, with a lateral shift of one or two repeat motifs, one $\alpha$-helix would bridge the gap to the next coiled-coil via the leucine zipper motif.
The fibril chain is then stabilized by the same hydrophobic contacts that stabilize the coiled-coil
(structures A and B in FIG.~\ref{fig:Lp}). 
Because of the overlap, we expect a stiffer fibril than in the aggregate with zero overlap.
To test this, we measured the fibril chain flexibility in models with different lateral shifts and compared the computational values to the results of SANS experiments \cite{Hellmund2021}.
Only structures with no or small lateral shifts were stable in our simulations (structure A, B, and C in Fig.~\ref{fig:Lp} B). 
We also include a continuous coiled-coil with the same peptide sequence for comparison
(structure CC in Fig.~\ref{fig:Lp} B).

The persistence length $L_P$, or chain decorrelation length, is a measure for the flexibility of a chain \cite{Boal2012}\cite{Li2010}. 
When $l$ is the position along the chain, one compares the local chain orientation at $l$, $\vec{e}(l)$, to the chain orientation $\vec{e}(l+ \Delta l)$  at $\Delta l$ further down the chain. 
$\vec{e}(l)$ and $\vec{e}(l+ \Delta l)$ are unit vectors.
$L_P$ is the length at which the chain orientation at $l+ \Delta l$ is fully uncorrelated from the chain orientation at $l$ (see section \ref{sec:methods}).

\begin{figure}[h!]
    \centering
    \includegraphics[width=0.9\columnwidth]{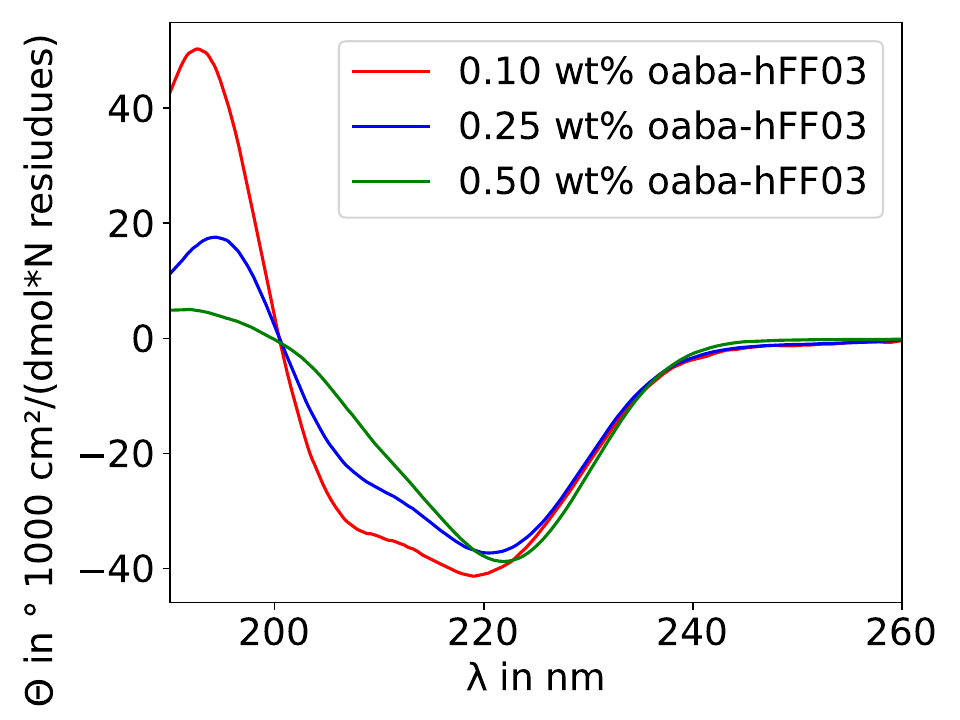}
    \caption{Circular dichroism spectra oaba-hFF03 in water at pH~7.4,  $T=37^{\circ}$C and concentration in mass percentage.}
    \label{fig:CD}
\end{figure}

To calculate $L_P$ we discretize the coiled-coil chain.
A discretization along the peptide backbone (as pre-implemented in programs that calculate peptide chain flexibility) returns the flexibility of the $\alpha$-helix within the coiled-coil, not the flexibility of the coiled-coil fibril chain.
Instead we define a new virtual chain of beads following the core of the coiled-coil, shown in FIG.~\ref{fig:Lp} A.
We define a bead for each leucine-leucine zipper pair, where the bead position is the midpoint between the leucine $C_{\alpha}$ atoms which places the beads at the center of the coiled-coil. 
FIG.~\ref{fig:Lp}.C compares the persistence lengths of our model systems to the experimental persistence length of oaba-hFF03 of 9 to 14~nm \cite{Hellmund2021}.
The continuous coiled-coil (CC), is very stiff with a persistence length of roughly 290 nm, which is in line with previously reported persistence lengths of peptide coiled-coils\cite{Li2010}. 
Interrupting the chain and introducing lateral shifts leads to more flexible fibril chains (A, B and C). 
However, the systems with overlapping peptide strands have persistence lengths that are much larger than the experimental value. 
System C, in which we manually aligned coiled-coil dimers with zero lateral shift into a straight fibril, is still stiffer than the experimental value, but the estimate shows a large standard deviation. 
Closer inspection of the simulations showed that the fibril chain started to fluctuate, leading to kinks at the interface between adjacent coiled-coil dimers, causing the large standard deviation. 
More specifically, the starting structure of the coiled-coil-dimer fibril was cut from a continues coiled-coil with the same heptad repeat unit, generating a very tight dimer-dimer interface. 
In half of the simulations these interfaces loosened.
This also implies that the manually aligned, very straight fibril chain conformation is not at a free-energy minimum, and the conformational equilibrium is more dynamic. 
To test this, we simulated randomly placed coiled-coil dimers of no-hFF03 in water. 
We observed that these coiled-coil dimers rapidly self-assemble into fibril chains within nanoseconds. 
However, these fibrils are highly dynamic and continuously break and reassemble.
We repeated these simulations with oaba-hFF03 and paba-hFF03, which showed the same behaviour.
The persistence lengths of these self-assembled fibril chains are shown as green ranges in FIG.~\ref{fig:Lp}.C and agree well with the experimental value.
We conclude from Fig.~\ref{fig:Lp} that coiled-coil assemblies with non-zero lateral shift are too stiff compared to the experimental persistence length.
Therefore, fibril formation via the leucine zipper motif can be ruled out.
Instead, coiled-coil dimers with zero later shift (model C) self-assemble into highly dynamics fibril chains with a persistence length of about 10~nm.
These fibrils are likely stabilized by hydrogen bonds and salt-bridges between the N- and C-termini of adjacent coiled-coils.
This two-step mechanism (coiled-coil formation, followed by self-assembly into fibrils) is supported by circular dichroism (CD) spectroscopy (FIG.~\ref{fig:CD}).
At low concentrations, the characteristic signals for $\alpha$-helical structures, namely an ellipticity maximum at 195 nm and two minima at 208 nm and 222 nm, respectively, are clearly present. 
The ellipticity minimum at 222 nm is of higher intensity than the minimum at 208 nm, indicating the formation of more ordered structures \cite{pandya2000sticky}.
The intensity ratio of these two minima increases as the concentration is increased, indicating that at higher concentrations more highly-ordered structures are formed. 
A possible mechanistic interpretation is that, at low concentrations, individual coiled-coils are present, which only self-assemble into fibrils if a certain concentration threshold is passed. 

\subsection{Influence of the chromophore on the fibril stability}
\begin{figure}[h!]
        \centering
    \includegraphics[width=6cm]{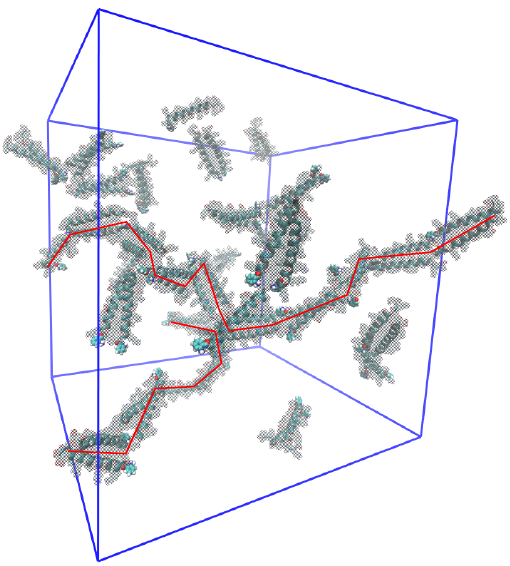}
         \caption{Snapshot of oaba-hFF03 in explicit water after 50~ns simulation (4 wt\%, starting structure with randomly placed coiled coils). Red lines highlight self assembled oligomers.
         }
         \label{fig:SelfAssembly}
\end{figure}

\begin{figure}[h!]
        \centering
    \includegraphics[width=\columnwidth]{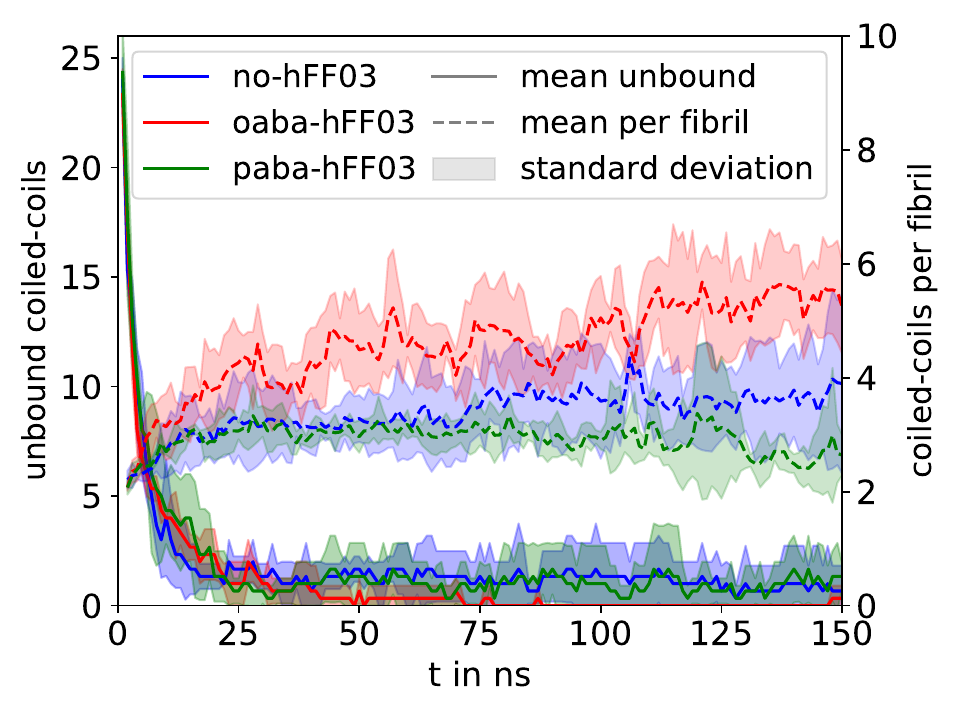}
         \caption{Time-series of the number of unbound coiled-coil dimers and number of coiled-coil dimers per oligomer chain. 
         For each system, we conducted three MD simulations.
         Average and standard deviation where calculated as running averages with a block size of 1~ns and then averaged over the three separate trajectories for each systems.}
         \label{fig:timeseries}
\end{figure}

Next, we investigated the influence of the chromophore on the structure and stability of the self-assembled fibril chains. 
For all three systems, we simulated 32 randomly placed coiled-coil dimers in a water box of 20x20x20~nm$^3$ for 150~ns. For each system, three seperate simulations with random start configurations where performed.
These systems corresponds to a polymer mass fraction of 4~\%.
A snapshot of the oaba-hFF03 simulation after 50~ns is shown in FIG.~\ref{fig:SelfAssembly} (See SI FIG. 2 for a snapshot of no-hFF03 and paba-hFF03).\\
The coiled-coil dimers rapidly form oligomers during the first few nano seconds. These oligomer chains then elongate. The elongated oligomers are highly dynamic and keep rearranging on a timescale of nanoseconds.  The progress of the oligomer formation is illustrated in FIG.~\ref{fig:timeseries}. The number of coiled-coil dimers, which are not part of an oligomer (solid lines), rapidly falls in the first few nanoseconds and reaches a plateau after about 50~ns. While for no-hFF03 and paba-hFF03 we find 1-2 unbound coiled-coil dimers per simulation box, all coiled-coil dimers in the oaba-hFF03 simulations are bound in oligomers.
As the number of unbound coiled-coils decreases, the average size of the oligmers increases (dashed lines).
It reaches a plateau of 2 to 3 coiled-coils per oligmer for paba-hFF03. 
In  no-hFF03 and oaba-hFF03, we observe a slight drift to longer oligomers throughout the simulation. 

We use the last 50 ns of the simulations to extract statistics on the oligomer sizes (Fig.~\ref{fig:fibreStability} A) and life-times (Fig.~\ref{fig:fibreStability} B).
The oligmer size distribution (Fig.~\ref{fig:fibreStability}.A) shows sizeable statistical uncertainties, which indicates that simulations are not fully converged, yet. 
Nonetheless, some trends are evident.
With a maximum probability at four coiled-coil dimers per oligomer, oaba-hFF03 forms longer oligomer chains than the other two system, whose distributions peak at two coiled-coils dimers per oligomer. 
Additionally, oaba-hFF03 forms the longest chains with up to 12 coiled-coils dimers per oligomer. 

Fig.~\ref{fig:fibreStability}.B shows the life-time distribution of the coiled-coil interactions. 
Since the coiled-coils rearrange on the nanosecond timescale, the statistical uncertainty in this distribution is much smaller than in Fig.~\ref{fig:fibreStability} A. 
The fast dynamics is also evident from the fast decay of the life-time distribution (solid line): most coiled-coil dimer interactions only last a few 100 picsosecond, and almost all of them are broken within the first nanosecond. 
The dashed lines show the cumulative distributions, which reveal differences between the systems on longer timescales. 
Both oaba-hFF03 and no-hFF03 exhibit coiled-coil dimer interactions which last 50 ns and probably longer, whereas in paba-hFF03 virtually every interaction is broken within 10 ns.
The mean lifetimes for a coiled-coil dimer interaction in oaba-hFF03 and no-hFF03 are 1.3 ~ns (standard deviation 6.6~ns) and 1.3~ns (standard deviation 6.4~ns), respectively. 
By contrast, the mean lifetime in paba-hFF03 is only 0.4~ns (standard deviation 1.5~ns).
(The standard deviation is larger than the mean, because the distributions are highly skewed. Negative connection times are not possible.)
Note however that the disruption of an individual coiled-coil dimer interaction, does not mean that the oligomer chain falls apart. 
Rather, the two stubs quickly reassemble with the same or other nearby oligomers into new oligomer chains.
Overall, these data show that the presence of the aba chromophore influences the size and stability of the coiled-coil oligomers. 
Strikingly, even the position of the amino group in the aba 
chromophore influences the size and stability of the coiled-coil oligomers.

\subsection{Structural analysis of the coiled-coil interface}
\begin{figure}
    \centering
    \includegraphics[width=\columnwidth]{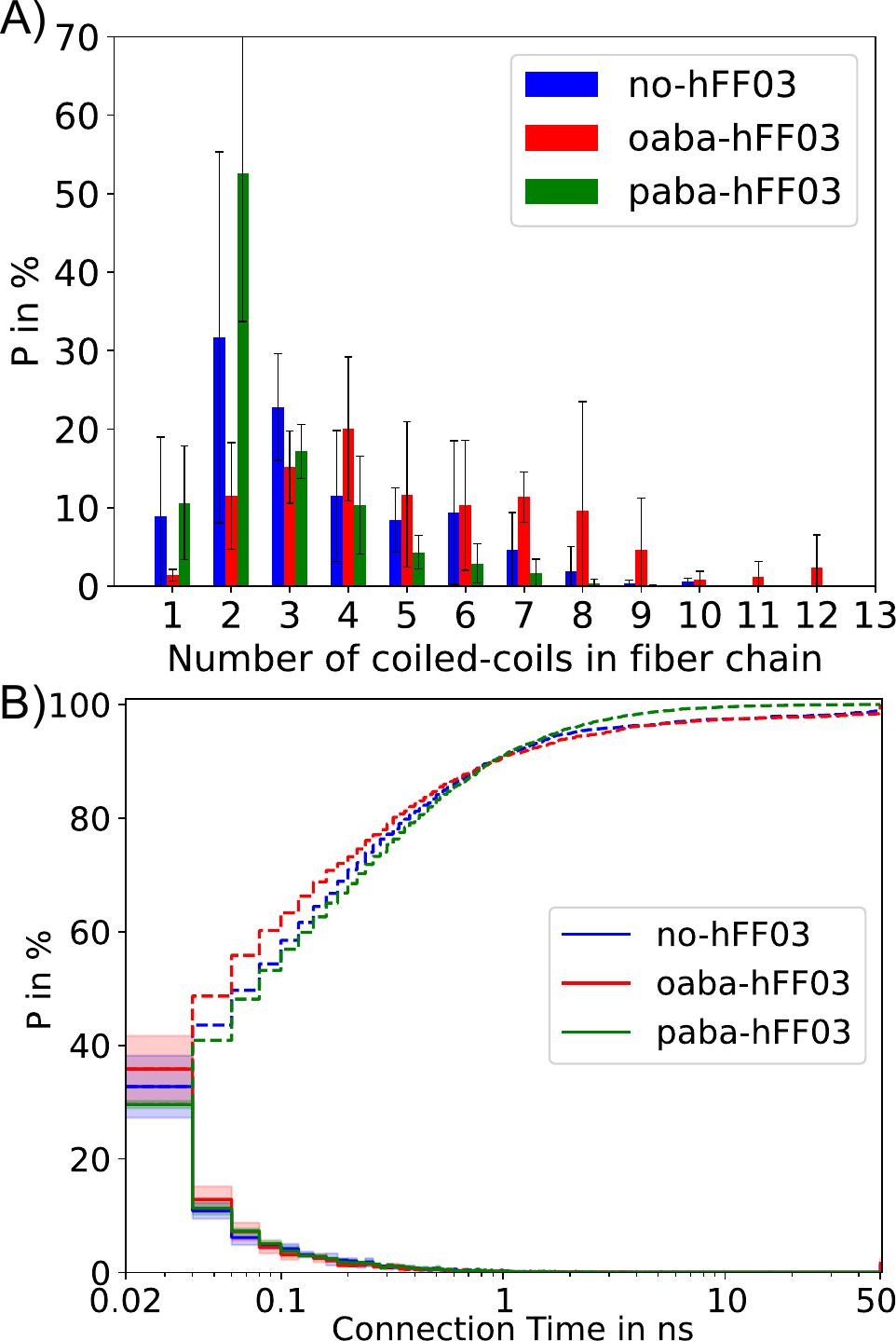}\\
    \caption{
    \textbf{A)} Oligomer size distribution between 100 and 150 ns of the MD simulation, averaged over three independent simulations for each system. Error bars indicate the standard deviation. 
    \textbf{B)} Residence time distribution of the coiled-coil dimer interactions (solid line) and corresponding cumulative distribution (dashed line) between  100 and 150 ns of the MD simulation. oaba-hFF03 exhibits coiled-coil dimer interactions that do not break in this time interval.}
    \label{fig:fibreStability}
\end{figure}
\begin{figure*}
\includegraphics[width=14cm]{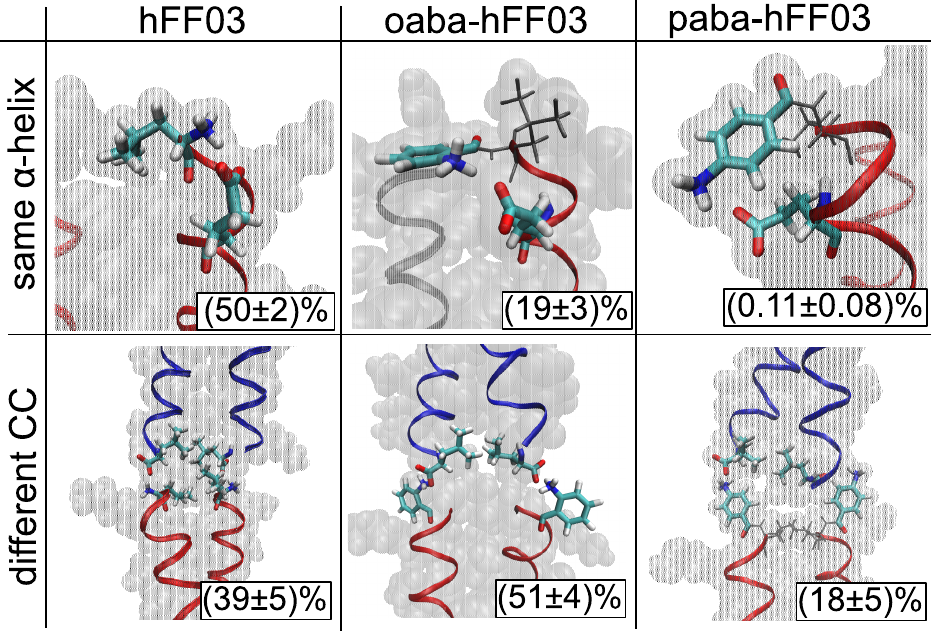}
    \caption{Influence of the chromophore label aba on the coiled-coil dimer interaction. 
    Upper row: salt bridges of the N-terminus within the same $\alpha$-helix, 
    lower row: salt bridges of the N-terminus to another coiled-coil dimer.
}
    \label{fig:comparison}
\end{figure*}

To elucidate the structural origin for the variation in the oligomer sizes and life-times, we analyzed the hydrogen bonds and salt bridges in the coiled-coil dimer interface. 
In no-hFF03, the interface between two coiled coils is predominantly stabilized by a salt bridge between the positively charged amino group of the N-terminus and the negatively charged carboxyl group of the the C-terminus (Fig.~\ref{fig:comparison}).
Additionally, the nearby lysine K2 can take the role of the N-terminus, and the nearby glutamic acid E25 in the other coil-coil can take the role of the C-terminus, such that in total the interface is stabilized by several fluctuating salt-bridges.
In oaba-hFF03 and paba-hFF03, the aba-group is covalently attached to the N-terminus, and the amino group of the aba group replaces the N-terminal aminogroup in the interface. 
We model the amino group in aba as protonated and positively charged and observe salt bridges from the amino group to the C-terminus and the glutamic acid in the adjacent coiled-coil.
The salt bridge between coiled-coil dimers competes with an intramolecular hydrogen bond (i.e.~within the same peptide chain) between the protonated amino group (N-terminus or aba) and the carboxyl group of the nearest glutamic acid: E4  (Fig.~\ref{fig:comparison}).

While we observe the same types of salt bridges in all three systems, the relative populations vary drastically across the systems.  (See SI FIG. 3-5 for all salt bridges.)
In no-hFF03 the protonated amino group forms an intra-molecular salt bridge in 50\% of all frames and a salt bridge to an adjacent coiled-coil in 39\% of all frames. 
Interactions with the surrounding water are observed in only 11\% of all frames. 
In oaba-hFF03, this equilibrium shifts in favor of the inter-coiled-coil salt bridge which is now populated to 51\%. 
The population of the intramolecular salt-bridge is decreased to 19\%, and interactions with water slightly go up to 30 \%.
In paba-hFF03, the situation is quite different from the other two systems. The amino group most frequently interacts with the surrounding water, and engages only in 18\% of all frames in an inter-coiled-coil salt bridge. 
The intra-molecular salt-bridge is almost never populated (0.1\%) 
See Tab.~\ref{tab:NTerm}.

\begin{table}[]
    \centering
    \begin{tabular}{l|c|c|c|c}
                & \multicolumn{4}{c}{N-terminal salt bridge to:  / in \%} \\
         & same $\alpha$-helix & same CC& different CC & solvent \\\hline
         no-hFF03 &50 $\pm$ 2 & 0.7 $\pm$ 0.3 & 39 $\pm$ 5 & 11 $\pm$ 7\\
         oaba-hFF03 & 19 $\pm$ 3 & 0.8 $\pm$ 0.5 & 51 $\pm$ 4 & 30 $\pm$ 6 \\
         paba-hFF03 & 0.11 $\pm$ 0.08 & 0.14 $\pm$ 0.12 & 18 $\pm$ 5 & 82 $\pm$ 5
    \end{tabular}
    \caption{Salt bridge population of the N-terminal amino group. See FIG.~\ref{fig:comparison} for examples of the different salt bridges. Populations are calculated over all salt bridges in a given category and all $\alpha$-helices in the simulation box. Mean and standard deviation are calculated from the three independent simulations for each system.}
    \label{tab:NTerm}
\end{table}

The steric arrangement of the salt-bridge seems to be the cause for this shift in the salt-bridge populations.
By moving the protonated aminogroup from the C-terminus to the aba-chromophore, one moves it away from the carboxyl group of E4, thus weakening the intramolecular salt bridge. 
Because in oaba, the aminogroup is in \emph{ortho}-position the oaba-group can be oriented such that the aminogroup points towards the E4. 
This is not possible if the aminogroup is in \emph{para}-position, and hence the intramolecular salt bridge is not formed in paba-hFF03. 
On the other hand, moving the protonated aminogroup to the aba group makes the coiled-coil interface less crowded than in no-hFF03. 
Fig.~\ref{fig:comparison} shows that in oaba-hFF03 the salt bridge can turn outside towards the solvent, leaving enough space for the two L1 and the two L26 residues to orient themselves in the hydrophobic center of the interface. 
Finally in paba-hFF03, the protonated amino group points towards the solvent, and it is difficult to find a conformation in which both salt bridges are formed across the coiled-coil interface. 
Two other effects contribute to the stability of the coiled-coil interface. First, as mentioned above, the salt bridge between the protonated amino group and the deprotonated carboxyl group of the C-terminus in the adjacent coiled-coil can be replaced by salt bridges involving K2 and E25. 
However, the stability of these salt bridges follows the same trend as the dominant salt bridge. 
Second, one can speculate that hydrophobic effects will play a role. 
The aba-group certainly makes the N-terminus more hydrophobic. 
On the other hand, the charges at the N-terminus of no-hFF03 are often capped by the intramolecular salt-bridge, which also generates a relatively hydrophobic N-terminus. 
Quantifying these hydrophobic effects is difficult.

In summary, the steric ease with which the salt bridge between the protonated amino groups at the N-terminal ends at one coiled-coil and the C-terminal carboxyl group at the adjacent coiled-coil can be formed determines how stable the coiled-coil interface is.
The stability of this salt-bridge (Fig.~\ref{fig:comparison} ) is directly correlated to the oligomer size distribution of the three substances (Fig.~\ref{fig:fibreStability}.A). 
However, the stability of the salt-bridge and the oligomer size distribution (on the timescale of our simulations) do not explain the differences in the mechanical properties of the hFF03-hydrogels  (Fig.~\ref{fig:Rheo}).

%---------------------------------------
%   W A T E R   R E T E N T I O N 
%---------------------------------------
\subsection{Water retention in the hFF03 hydrogels}
To get a better understanding of how the coiled-coils influence the structure and dynamics of the nearby water molecules, we analyzed the water density and local diffusion constant in the hydration shell of a no-hFF03 coiled-coil. 
The self diffusion coefficient of a particle $D$ is a  transport property which is usually calculated as the slope of the mean-square displacement versus time. 
Because the particle is allowed to diffuse away from its initial position during this calculation, this estimator is not suited to calculate local self-diffusion constants. 
Instead one can use the Green-Kubo relations \cite{Kubo1966} to express this transport property as the particle's velocity auto-correlation function (VACF):
\begin{equation}
    D_{\mathrm{self}}= \int_{\tau = 0}^{\infty} \langle v(0)v(\tau)\rangle d\tau
    \label{eq:Diff}
\end{equation}
where $v(0)$ is the particle's velocity at time $t=0$, $v(\tau)$ is its velocity at time $t=\tau$, and the VACF $\langle v(0)v(\tau) \rangle$ is the ensemble average of $v(0)v(\tau)$. 
The ensemble averages for different lag times $\tau$ are integrated from $\tau =0$ to $\tau = \infty$ to obtain the diffusion constant. 
Of course, for long lag times $\tau$, the particle diffuses away from its initial position.
But in water, the VACF levels off to zero at around 800~fs and the integral converges at an integration limit of a few picoseconds.
This is shorter than the time scale of structural rearrangements in the solvation shell, which usually take tens of picoseconds. 
Thus, eq.~\ref{eq:Diff} indeed allows us to define a spatially resolved diffusion constant.
Eq.~\ref{eq:Diff} is calculated using the the Fast Fourier Transformation algorithm. See section \ref{sec:methods}.
Note that the self diffusion coefficient of bulk TIP3P water of $D_{\mathrm{self}} =6.78\cdot 10^{-5} \, \mathrm{cm}^2\mathrm{/s}$  obtained from our calculations is slightly higher than previously reported values calculated using the mean-square displacement \cite{braun2014transport, mark2001structure} with values around $D_{\mathrm{self}} = 5.8 \cdot 10^{-5}\, \mathrm{cm}^2\mathrm{/s}$. However the water density is also slightly lower for our simulations.
While it is known that TiP3P water has a higher self diffusion coefficient than the experimentally obtained value of $D_{\mathrm{self}} = 2.3 \cdot 10^{-5}\, \mathrm{cm}^2\mathrm{/s}$ at 300~K this should not detract from the results\cite{harris1980pressure}.

\begin{figure}[h]
    \includegraphics[width=\columnwidth]{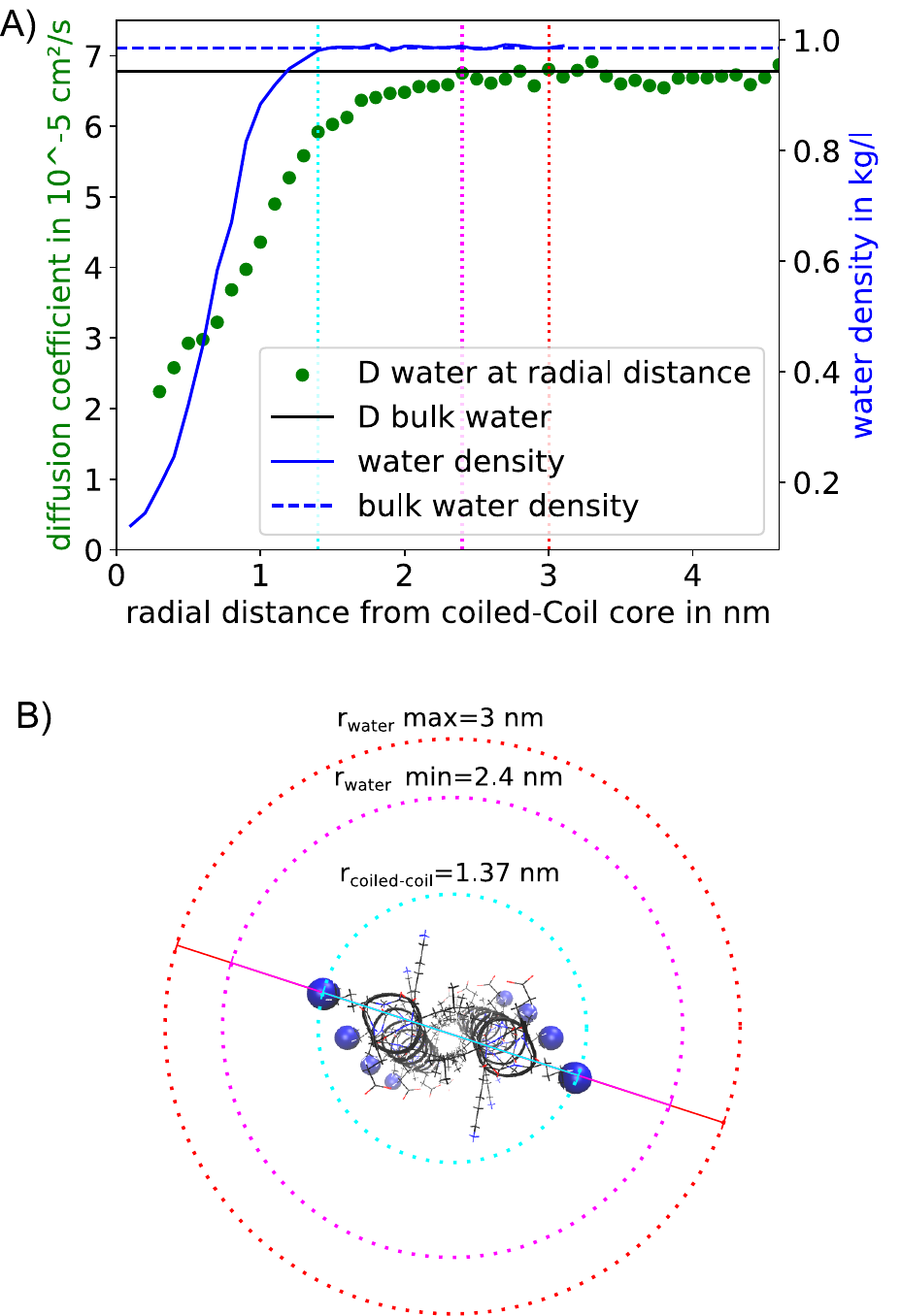}
    \caption{\textbf{A}: local density and diffusion constant of water in the vicinity of hFF03. The distance is the radial distance of the water molecule to the center of the coiled-coil. The bulk values were calculated from MD simulation box of pure water.
    Vertical lines show radii at which bulk properties are restored: density (cyan), diffusion constant, lower boundary (pink), diffusion constant, upper boundary (red).
    \textbf{B}: visualisation of the radii around a coiled-coil dimer. The radius at which density of bulk water is reached coincides with the coiled-coil dimer radius (cyan).}
    \label{fig:Diff}
\end{figure}

The solid blue line in Fig.~\ref{fig:Diff}.~A shows the local water density around the coiled-coil dimer. 
It increases from zero at the center of the coiled-coil to the density of bulk water. 
The density curve levels off at $r\approx 1.37 \,\mathrm{nm}$, which is very close to the coiled-coil radius that we determined in 
Fig.~\ref{fig:Oligomerisation}.
Because water molecules can penetrate in between the side chains, the water density does not immediately drop to zero at this radius but slowly decreases. 
However, the influence of the coiled-coil dimer reaches beyond  $r\approx 1.37 \,\mathrm{nm}$. 
This can be seen from the local diffusion constant, which only reaches the bulk diffusivity at between $r=2.4\,\mathrm{nm}$
and $r=3.0\,\mathrm{nm}$.
That is, within the solvation shell from 1.4~nm to 2.4~nm, rearrangements of the water structure are slower than in bulk water. 

By comparing the volume of at the coiled-coil dimer including its hydration shell to the volume of the simulation box, we can calculate how much bulk water remains in the system.  
We approximate the shape of the coiled-coil dimer including its solvation shell as a cylinder whose volume is
$V_{\mathrm{cylinder}}= \pi r^2 h$, where $r$ is the radius of the cylinder and $h$ is its height. 
We set $r=2.4\,\mathrm{nm}$ (radius of the hydration shell) and $h = 4.0\,\mathrm{nm}$ (length of the no-hFF03 coiled-coil dimer), which yields a volume of $V_{\mathrm{cc}}= 72.4\, \mathrm{nm}^3$. 
Our simulation box has a volume of  $V_{\mathrm{box}} = 20 \times 20 \times 20 \mathrm{nm}^3 = 8000\,\mathrm{nm}^3$ 
At a concentration of 4 wt\%, it contains 32 coiled-coil dimers. 
Thus, the volume fraction of bulk water is $1 - 32\cdot V_{\mathrm{cc}}/V_{\mathrm{box}} = 0.710$, i.e. 71\%.
If we assume that the hydration shell reaches up to $r=3.0\, \mathrm{nm}$, the bulk water content decreases to 55\%. 
In either case, at these high concentrations, a hFF03-hydrogel consists to a substantial part of coiled-coil peptides and water bound to these peptides. 
However, Fig.~\ref{fig:CD} shows that fibril formation sets in at much lower concentrations. 
In Fig.~\ref{fig:Rheo}, the mechanical properties of the hFF03-hydrogel have been measured at 0.5 wt\%.
At this concentration, the bulk water content ranges between 96\% (for $r=2.4\, \mathrm{nm}$) and 94\% (for $r=3.0\, \mathrm{nm}$).
That is, almost everywhere in the sample the water moves unhindered. 
Retention of water close to the coiled-coil fibrils therefore seems to contribute little, if at all, to the observed hydrogelation. 

%--------------------------------------------------------
%                   Discussion
%------------------------------------------------------
\section{Conclusion}
We developed an atomistic model of the self-assembled peptide hydrogel hFF03. 
The peptides within this structure form $\alpha$-helical coiled-coils with zero lateral shift. 
These coiled-coil dimers self-assemble into oligomers on the nanosecond timescale by forming salt-bridges between C and N-terminus of neighboring coiled-coil dimers.
Our model aligns well with previously published small-angle neutron scattering (SANS) data and circular dichroism (CD) spectra presented in this study.
Specifically, the oligomers match the experimental diameter and persistence lengths.
Our model refutes the possibility of a sticky-end assembly, where hydrophobic contacts between leucine residues would stabilize the oligomers. 
The chromophore aminobenzoic acid, which was originally added at the N-terminus as an UV-VIS marker, has a significant impact on the structure and dynamics of the coiled-coil dimer interface. 
Altering the position of the amino group from ortho- to para-position in aba shifts the equilibrium between salt bridges and hydrogen bonds with the surrounding water, subsequently influencing the sizes and stability of the oligomers.
The striking consequence of this is that the chromophore label controls the rheological properties of hFF03 hydrogels. 
In fact, the presence of the chromophore label is crucial for hydrogel formation, as no-hFF03 does not form a hydrogel.
oaba-hFF03, which generates the longest and most stable oligomers of the three variants, also forms the most stable hydrogel at the macroscopic level. In contrast, paba-hFF03 forms shorter oligomers and also a softer gel with a markedly shorter structural relaxation time
Our model does not fully account for the macroscopic results, and especially the oligomer size distribution does not fully correlate with the rheological data. 
How long range interactions necessary for elastic component of the viscoelastic properties arise from the rapid coiled-coil rearrangements observed in our simulations, is not yet explained.
The largest difference between simulations and oscillatory shear experiments is the timescale. 
With an aggregated simulation time of 450~ns per system, we have here probed the dynamics of hFF03-hydrogels on timescales from $10^{-12}$ to $10^{-7}$ seconds (THz to 0.1~MHz). 
By contrast, the oscillatory shear experiments probed the dynamics on timescales from $10^{-2}$ to 10 seconds (10~mHz to 10~Hz), i.e., they probe the structural relaxation time of the complete system.
In contrast, the simulations only probe the first basic elementary step involved in the process of gelation, which is the coiled-coil dimer interaction. The whole rheological relaxation process is, of course, that of the whole system of many such cross-links. In that respect the situation is similar to that of hydrophobically cross-linked hydrogels, where the individual hydrophobic sticker has a lifetime of µs while the structural relaxation time will be in the range of many seconds, both scaling with the hydrophobicity of the sticker\cite{MaloDeMolina2012}.
Of course, the intermediate time-range can be interesting for a further understanding of the relation between rheological properties and mesoscopic structure. To narrow the timescale gap between simulation and experiment, one can prolong the atomistic simulations to cover timescales of $10^{-6}$ up to $10^{-4}$ seconds (1 MHz to 0.1 kHz). With coarse-grained simulations even longer timescales are accessible. 
% 
%On the experimental side, high-frequency rheometry gives access to the mechanical response of the hydrogel on a timescale of $10^{-5}$ to $10^{-1}$ seconds (100~kHz to 0.1~Hz) \cite{KIT}
%
Other concerns are the limited size of the simulation box, which might introduce a spurious periodicity in the system, and the water model used. 
Using a 4-site water model \cite{abascal2005general} or a polarizable water model \cite{liu2019implementation}, will yield a more realistic representation of the diffusive dynamics of the coiled-coil dimers and of the water structure in the solvation shell of the peptides.
Nonetheless, our study reveals that modifying the chromophore label is a synthetically simple strategy to shape the interactions between coiled-coil dimers and to thereby tune the viscoelastic properties of the peptide hydrogel.
By adding a chromophore label, a hydrophobic group is introduced at the N-terminus and the solvent exposed amino group is shifted away from the peptide backbone.
These two effects counterbalance each other. 
The presence of the hydrophobic group increases the hydrophobicity of the N-terminus, while the shift of the amino group weakens the intramolecular salt-bridge, thus increasing the solvent exposure of the amino group and a nearby glutamate residue.
The third effect of the chromophore label is sterically changing the salt bridge network that stabilizes the interface between two coiled-coil dimers.
By shifting the amino group away from the peptide backbone, the interface is sterically less crowded which increases oligomer size and stability in oaba-hFF03. 
However, when placing the amino group in para-position the amino group cannot as easily be oriented towards the interface, and as a consequence oligomer size and stability are lower in paba-hFF03 than in oaba-hFF03. 
In conclusion, the three parameters of the chromophore label - size of the aromatic system, distance between carboxyl group and aminogroup, orientation of the amino group relative to the carboxyl group -
open up a design strategy to control the viscoelastic properties of hFF03 peptide hydrogels.
\subsection{Acknowledgements}
Funded by the Deutsche Forschungsgemeinschaft (DFG, German Research Foundation) – SFB 1449 – 431232613; sub-projects C02 and A02
The authors would like to thank the HPC Service of ZEDAT, Freie Universität Berlin, for computing time\cite{Bennett2020}.

%\printbibliography
%%%%%%%%%%%%%%%%%%%%%%%%%%%%%%%%%%%%%%%%%%%
%
%   R E F E R E N C E S
%
%%%%%%%%%%%%%%%%%%%%%%%%%%%%%%%%%%%%%%%%%%%
%\newpage
\section{References}
\bibliography{bib}% Produces the bibliography via BibTeX.

\appendix
\section{Additional Figures}

\begin{figure}[h!]
    \centering
    \includegraphics[width=8cm]{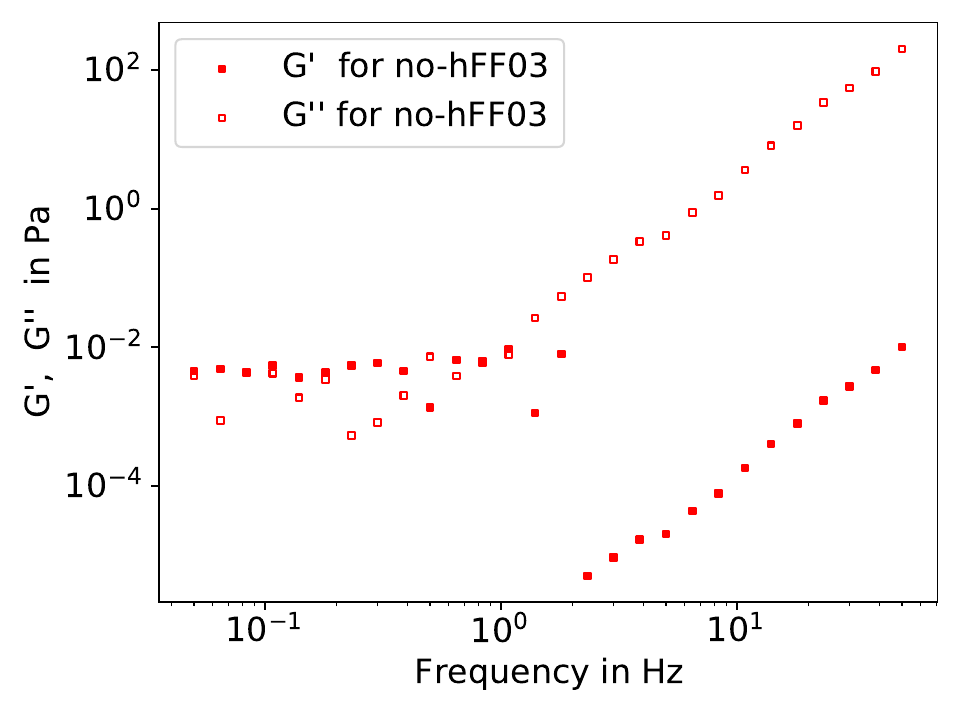}
    \caption{Frequency sweep measured through oscillatory shear experiment of no-hFF03. The shown features are a result of measurement artefacts and can be attributed to effects of instrument and fluid inertia.}%\cite{Lauger2016}
    \label{fig:Rheo2}
\end{figure}

\begin{figure}[h!]
\includegraphics[width=8cm]{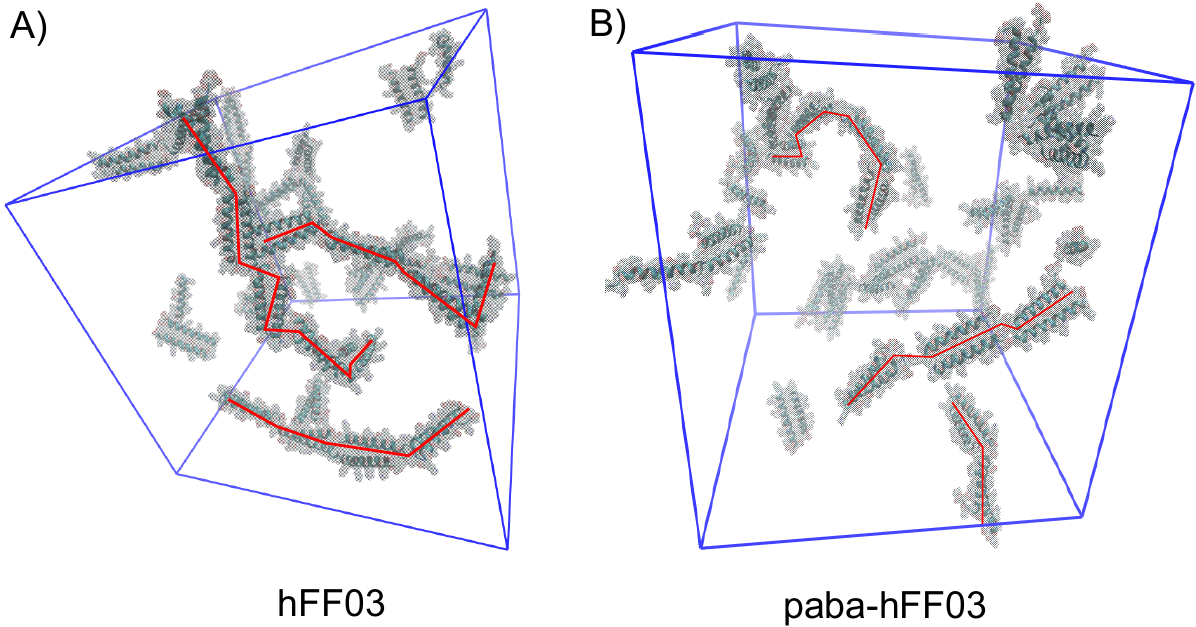}
    \caption{
    Snapshots of no-hFF03 (A) and paba-hFF03 (B) in explicit water after 50 ns simulation (4 wt\%, starting structure with randomly placed coiled-coils). Red lines highlight self assembled oligomers.
}
    \label{fig:NO-PAB}
\end{figure}

\begin{figure}[h!]
\includegraphics[width=8cm]{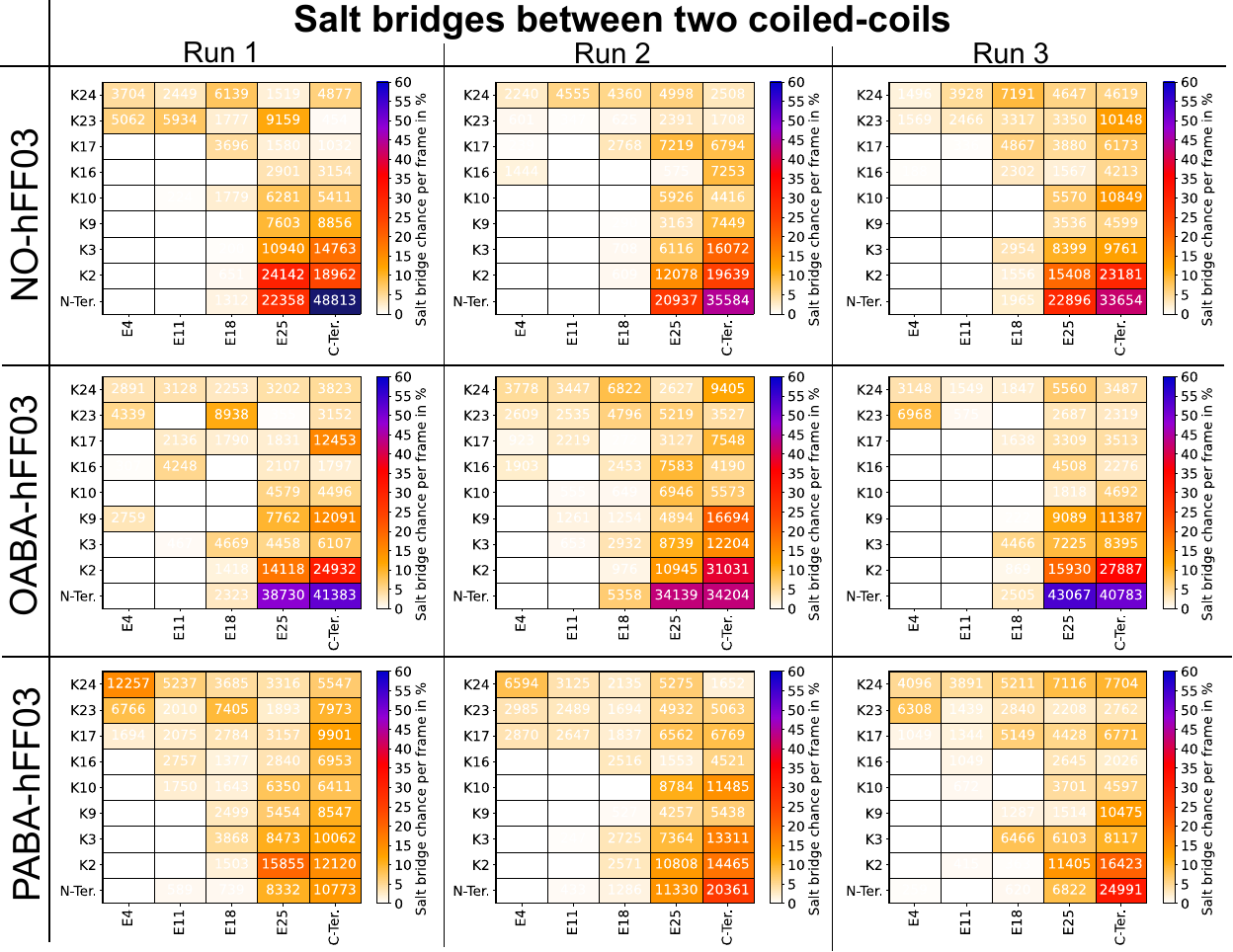}
    \caption{Populations of salt-bridges between of coiled-coil dimers. These salt-bridges stabilize the fibril. The numbers represents in the number of frames in which the salt bridge occurs in the time interval 100 ns to 150 ns of the MD simulation (total of 25.000 frames).}
    \label{fig:Inter_coil}
\end{figure}

\begin{figure}[h!]
\includegraphics[width=8cm]{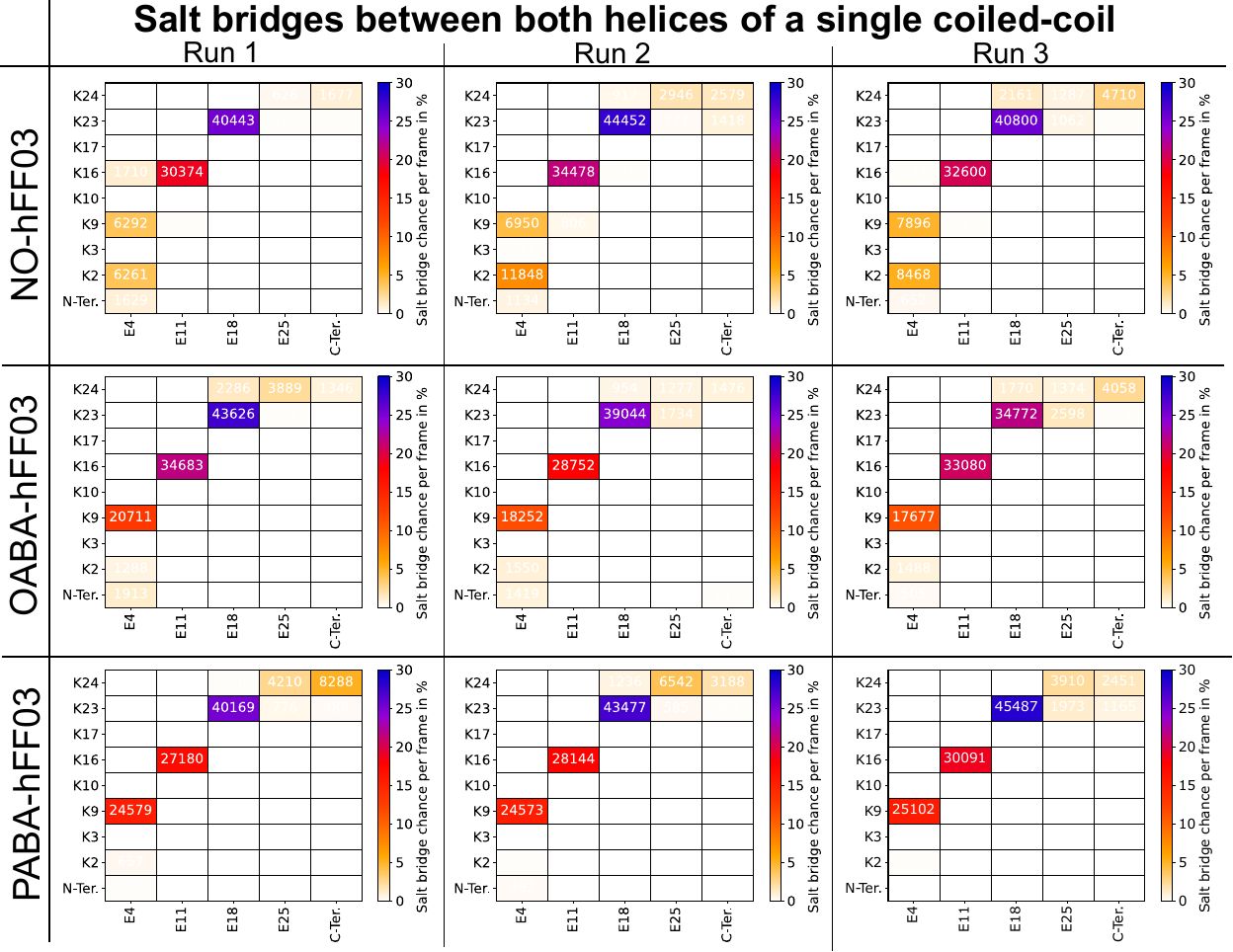}
    \caption{Populations of salt-bridges between of $\alpha$-helices within a coiled-coil dimer. These salt-bridges stabilize the dimer in addition to the leucine zipper motif. The numbers represents in the number of frames in which the salt bridge occurs in the time interaval 100 ns to 150 ns of the MD simulation (total of 25.000 frames).}
    \label{fig:Intra_coil}
\end{figure}

\begin{figure}[h!]
\includegraphics[width=8cm]{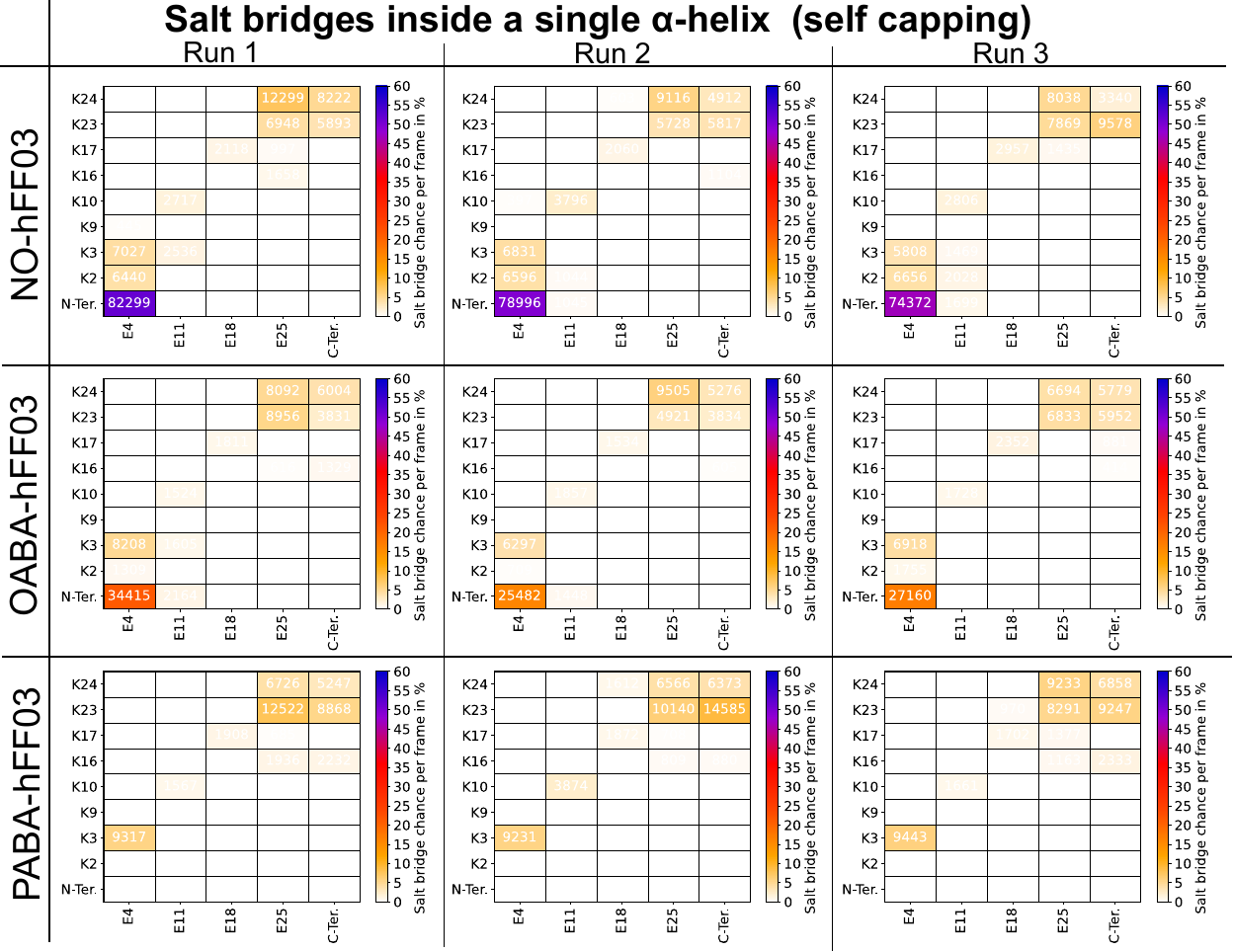}
    \caption{Populations of salt-bridges within a peptide chain. These salt-bridges stabilize the self-capping of the N-terminus. The numbers represents in the number of frames in which the salt bridge occurs in the time interaval 100 ns to 150 ns of the MD simulation (total of 25.000 frames).}
    \label{fig:Intra_helix}
\end{figure}

\end{document}